\documentstyle[sprocl,epsf]{article}

\def\be{\begin{equation}}
\def\ee{\end{equation}}
\def\bea{\begin{eqnarray}}
\def\eea{\end{eqnarray}}
\newcommand{\nn}{\nonumber}

\def\bfnabla{\mbox{\boldmath $\nabla$}}

\def\lQ{\Lambda_{\rm QCD}}

\def\als{\alpha_{\rm s}}
\def\siml{{\ \lower-1.2pt\vbox{\hbox{\rlap{$<$}\lower6pt\vbox{\hbox{$\sim$}}}}\ }}
\def\simg{{\ \lower-1.2pt\vbox{\hbox{\rlap{$>$}\lower6pt\vbox{\hbox{$\sim$}}}}\ }} 

\def\lla{\langle\!\langle}
\def\rra{\rangle\!\rangle}
\newcommand{\Appendix}[1]%
    {%
     \section{#1}%
      }
\def\siml{{\ \lower-1.2pt\vbox{\hbox{\rlap{$<$}\lower6pt\vbox{\hbox{$\sim$}}}}\ }} 
\def\bfnabla{\mbox{\boldmath $\nabla$}}

\newcommand{\AmS}{{\protect\the\textfont2 A\kern-.1667em\lower.5ex\hbox{M}\kern-.125emS}}

\begin{document}

\title{PROBING THE QCD VACUUM\footnote{Invited talk given at the 
XVIII Autumn School ``Topology of Strongly Correlated Systems'', Lisbon,
Portugal, 8-13 October, 2000.}}
\author{NORA BRAMBILLA}

\address{\it Institut f\"ur Theoretische Physik\\ 
Philosophenweg 16, Heidelberg  D-69120, Germany \\
E-mail: n.brambilla@thphys.uni-heidelberg.de
 \\ and Dipartimento di Fisica, U. Milano, Italy}
%\\
%Via Celoria 16, 20133 Milano, Italy}
 

\maketitle\abstracts{Heavy quark bound states are used as significative probes 
of the QCD vacuum and the mechanism of confinement.}

\section{Strong interaction and topological nontrivial configurations}
The focus of this school is on the ``Topology of strongly correlated 
systems'' in different areas ranging from QCD to condensed matter systems.
In all these fields, investigations were presented about nontrivial 
topological configurations. In QCD such configurations are believed 
to be related to the nontrivial structure of the QCD vacuum and to 
the confinement mechanism \cite{bak,pol,thof,rein,fab} on one hand and to the breaking of the chiral 
symmetry \cite{neu,thof,chir} on the other hand. However, the QCD dynamics is extremely complicate 
and in order to test our understanding of the topological configurations,
we need a systematic and under control way of parameterize the low energy 
physics in a frame that can still be simply related to the real experiments
and/or to the lattice experiments\cite{rev}. It is the aim of my talk to show how 
this is possible and in  which physical situations.

\section{Heavy Quark Bound States}
Heavy-quark bound systems are an ideal playground for theoretical 
ideas about strong QCD. In fact, while light quarks are highly 
``nonperturbative objects'', strongly interacting with the vacuum 
and with a  mass  almost entirely dominated by chiral symmetry breaking
(topological nontrivial) effects,
heavy quarks behave as ``external
 sources'' and thus as natural probes of the QCD vacuum. 
 Being heavy at least the quark mass scale $m$  can be 
treated perturbatively: $m \gg \Lambda_{\rm QCD}$. $\lQ$ has here always 
to be understood  as the scale at which nonperturbative effects become dominant.
Furthermore, since the characteristic difference $\Delta E$
between the energy levels 
 is $\Delta E \ll m$ inside such systems, they are 
nonrelativistic and can 
be described at leading order by the appropriate Schr\"odinger equation.
This gives origin, as usual in nonrelativistic bound state, to several 
dynamical scales below $m$:
the relative momentum $p\sim mv$ (the inverse of which 
gives the characteristic spatial size $r$  of the system), the energy $E \sim mv^2$ (the inverse 
of which gives the characteristic time $T$ of the system).
 Here, $v$ is the velocity 
of the heavy quark in the bound state. In QCD, $v$ is in general a function 
of the whole dynamics, perturbative and nonperturbative, but still remains a small parameter, $v\ll 1$,
and thus the scales mentioned above turn out to be  well separated and 
hierarchically ordered. It is well known that a technical (and quite 
hard) problem     in bound state calculation is originated 
by the fact that such scales get entangled \cite{nrrev}. 
Thus, even in QED it turns out 
to be very useful to disentangle such scales. In QCD this becomes a deeper 
and more conceptual issue, since we need to separate as much as possible
the perturbative contributions ('hard' physics at the high scale)
from the nonperturbative effects ('soft or ultrasoft' physics at the low
scale). {\it It is  on these last contributions that we will 
eventually be in the position to use our knowledge about 
topologically nontrivial configurations}.\par
I will show here that 
in  quarkonium physics the existence of  different dynamical scales  
besides $ \Lambda_{\rm QCD}$ 
\begin{itemize}
\item{makes the calculations  more complicate  from a technical point of view;}
\item{however, it makes also possible to  test  the non-perturbative  
nature of the  QCD vacuum  at different levels of deepness.}
\end{itemize}
In fact, heavy quark bound states provide a full set of probes that may 
explore different characteristic distances, from the small to the large distance regime,
 and different dynamical situations,
thus providing insights into   per\-tur\-ba\-ti\-ve/non-per\-tur\-ba\-ti\-ve fac\-to\-ri\-za\-tion  
mechanisms.
In Fig. 1 we show a phenomenological determination of the characteristic radius $r$
of heavy quarkonia against the phenomenological $Q\bar{Q}$ potential.
\begin{figure}
\vskip -0.15truecm
\makebox[0truecm]{\phantom b}
\put(110,0){\epsfxsize=5truecm \epsfbox{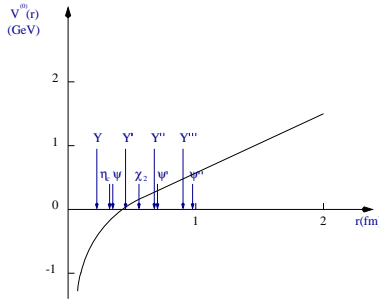}}
\caption{\it  $r_{Q\bar{Q}}$ vs Cornell potential (from Ref.$^8$).}
\vskip -0.35truecm
\end{figure}
It is apparent that heavy quarkonia  display a pattern of characteristic radii  
that extend from the perturbative region $r\ll \lQ $  to the
transition and the confinement region $r \gg \lQ$.
For ground state
  quarkonia (e.g. $\Upsilon(1S)$) $r \ll  1/\lQ$: 
 the scale $1/r$   is perturbative (this also implies that the binding 
potential is perturbative). For 
most quarkonia  $ r \sim  1/\lQ$:  the scale $1/r$ is nonperturbative.
\begin{figure}
\makebox[0truecm]{\phantom b}
\put(50,0){\epsfxsize=0.7truecm \epsfbox{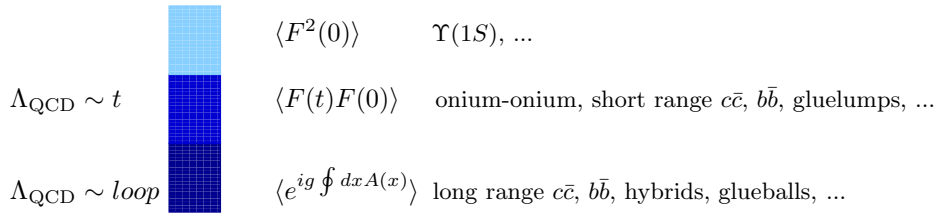}}
\put(90,65){$ \langle F^2(0) \rangle$ \quad~~~ {\small  $\Upsilon(1S)$, ... }} 
\put(90,40){$ \langle F(t)F(0) \rangle$ \quad 
{\small  onium-onium, short range $c\bar{c}$, $b\bar{b}$, gluelumps, ... }} 
%\put(153,50){\small  short range $c\bar{c}$, $b\bar{b}$}
%\put(153,35){\small  gluelumps,...} 
\put(90,5){$\langle \displaystyle e^{ig\oint dx A(x)}\rangle$}
\put(150,5){\small  long range  $c\bar{c}$, $b\bar{b}$, hybrids, glueballs, ... } 
%\put(150,-10){ hybrids, glueballs, ...}
\put(-10,40){$ \Lambda_{\rm QCD} \sim t$}
\put(-10,5){$ \Lambda_{\rm QCD} \sim loop$}
\caption{\it The more extended the physical object, the more sensitive 
to non-perturbative physics.}
\vskip -0.65truecm
\end{figure}
In order to be able to perform a systematic study, we need a clean 
and under control approach that allows us to  factorize and resum 
the perturbative physics at the appropriate scale, leaving us with the 
appropriate parameterization of the nonperturbative physics at the low 
energy scale. This goal is achieved by constructing an effective field theory (EFT).
 The existence of a hierarchy of energy scales in quarkonium systems allows 
the construction of EFT with less and less dynamical degrees of freedom but 
completely equivalent to QCD. This leads ultimately to 
 a field theory derived  {\it 
quantum mechanical  description of these systems}.
We call potential nonrelativistic QCD (pNRQCD) the corresponding EFT.
pNRQCD provides an unambiguous power counting in $v$ which determines which operators are relevant
at a given order of the expansion in $v$.
We refer to  \cite{pnrqcd,nrrev,pnrqcdrev} for details about pNRQCD. Here, we concentrate on the 
nonperturbative contributions: depending on the extension of the 'probe' systems
such contributions enter parameterized in a completely different way. This is precisely 
the interesting point that may allow us to understand something about the QCD vacuum structure.
In QCD we can never get completely rid of the nonperturbative effects.
However,  I will show that a pattern like the one represented in Fig 2 is 
realized 
in the heavy quark bound state nonperturbative dynamics. 
 If the spatial 
and temporal characteristic  scales of the bound system are perturbative (i.e. if 
$mv\gg mv^2 \gg \lQ$), then nonperturbative 
contributions arise only in the form of local gluon  condensates
 which contribute to the energy levels. When the characteristic 
temporal scale starts to be dominated by $\lQ$ ($mv^2 \sim \lQ$), then the nonperturbative contributions 
are encoded in nonlocal (time-dependent) gluon condensates. As soon as 
$1/\lQ$ falls between the spatial and the temporal scale 
($mv > \lQ > mv^2$), the nonperturbative 
nonlocal condensates enter the definition of the potential and give origin 
to short distance ($r< 1/\lQ$) nonoperturbative corrections.
At the end, when $\lQ$ is comparable to the spatial  size ($mv \sim \lQ$), the nonperturbative 
information is carried by bidimensional extended objects called Wilson loops.
The potential is  given in terms of them.
Quite interestingly, the Wilson loop turns out to be the order parameter 
of confinement (in gluodynamics) and allows us to put in direct connection topological configurations 
and the heavy quark phenomenology.\par 
As we see from Fig. 2, there are sufficient physical 
systems in nature to test all these dynamical situations. 
%However, sometimes the difficulty is to state  
%apriori in which situation a particular physical system lies.
%Thisis mainly due to the fact that the scales $mv$ and $mv^2$ are not always 
%so well defined nor so widely separated. 
%In these cases, the scales can only be fixed a posteriori, 
%i.e. by assuming that a particular situation holds and 
%by checking that the calculation is consistent with it. This is a general 
%feauture of EFT.

\section{Nonrelativistic quark  bound states  with a small characteristic \\ radius:
Condensates}
For heavy quark bound systems 
for which the following hierarchy $m \gg mv \gg mv^2 \simg \lQ$
holds, 
we can sequentially integrate out in perturbation theory both the hard scale $m$  and the soft 
scale $mv$. Then $v\sim \alpha_s$.
We denote by ${\bf R}\equiv {({\bf x}_1+{\bf x}_2})/2$ the center of mass of the $Q\bar{Q}$ 
system and 
by  ${\bf r\equiv {\bf x}_1 -{\bf x}_2}$ the relative distance. 
Then at the scale of the 'matching' $\mu^\prime$ 
($mv \gg \mu^\prime \gg mv^2, \lQ$) we have 
as still dynamical degrees of freedom: $Q\bar{Q}$ states  with energy of order of the next relevant 
scale, $O(\Lambda_{QCD},mv^2)$, momentum  of order $O(mv)$,  plus 
ultrasoft gluons $A_\mu({\bf R},t)$ with energy 
and momentum of order  $O(\lQ,mv^2)$. Therefore only very low energy gluons remain in the EFT 
as dynamical degrees of freedom: they are {\it multipole 
expanded}.  We find convenient to decompose the $Q\bar{Q}$ state 
into  a singlet $S({\bf R},{\bf r},t)$ and an octet $O({\bf R},{\bf r},t)$
with respect to  color transformations.
The pNRQCD Lagrangian is given 
at next to leading   order (NLO)  in the multipole expansion by\cite{pnrqcd}:
\begin{eqnarray}
& &\hspace{-4mm}
{\cal L}=
{\rm Tr} \Biggl\{ {\rm S}^\dagger \left( i\partial_0 - {{\bf p}^2\over m} 
- V_s(r) -\sum_{n=1} {V_s^{(n)}\over m^n} \right) {\rm S} 
+ {\rm O}^\dagger \left( iD_0 - {{\bf p}^2\over m} 
- V_o(r) - \sum_{n} {V_o^{(n)}\over m^n}  \right) {\rm O} \Biggr\}
\nonumber\\
& & \!\!\!\!\!\!\!\!\!\!
 + g V_A ( r) {\rm Tr} \left\{  {\rm O}^\dagger {\bf r} \cdot {\bf E} \,{\rm S}
+ {\rm S}^\dagger {\bf r} \cdot {\bf E} \,{\rm O} \right\} 
   + g {V_B (r) \over 2} {\rm Tr} \left\{  {\rm O}^\dagger {\bf r} \cdot {\bf E} \, {\rm O} 
+ {\rm O}^\dagger {\rm O} {\bf r} \cdot {\bf E}  \right\} -{1\over 4} F^a_{\mu\nu}
F^{\mu \nu a}.  
\label{pnrqcd0}
\end{eqnarray}
All the gauge fields in Eq. (\ref{pnrqcd0}) are evaluated 
in ${\bf R}$ and $t$. In particular ${\bf E} \equiv {\bf E}({\bf R},t)$ and 
$iD_0 {\rm O} \equiv i \partial_0 {\rm O} - g [A_0({\bf R},t),{\rm O}]$. 
The $V_j$ are potential matching coefficients: 
they are in the EFT  the correspondent to the  
usual notion of the potential. 
We call $V_s$ and $V_o$ the singlet and octet static matching potentials respectively.
The singlet sector of the Lagrangian gives rise to equations of motion of the 
Schr\"odinger type, while the last line of (\ref{pnrqcd0})
contains  retardation (or nonpotential) effects that 
start at the NLO in the multipole expansion. At this order the nonpotential
effects come from the singlet-octet and octet-octet
interaction mediated by an ultrasoft chromoelectric 
field.
Recalling that ${\bf r} \simeq 1/mv$ and that the operators count like the next relevant 
scale, $O(mv^2,\lQ)$, to the power of the dimension, it follows that  
each term in  the pNRQCD Lagrangian has a definite power counting.  
In the EFT language the potential is defined upon 
the integration of all the scales {\it up to the ultrasoft 
scale $mv^2$} and thus if  $\lQ \siml mv^2$
 {\it the potential is given by a pure perturbative expansion
in $\alpha_s$ at all orders}. 
The EFT gives us with (\ref{pnrqcd0})
a precise prescription to calculate both potential and nonpotential 
contributions to the energy levels and to count their relevance in power of $\alpha_s$. 
Nonpotential effects start only at order $\alpha_s^5 \ln \mu^\prime$.
At this order (three loops)
 the static singlet potential starts to become sensitive to the infrared 
physics via a logarithmic dependence on $\mu^\prime$\cite{pnrqcd}.
Such  dependence  is re-absorbed in a physical observable by non-potential 
contributions\cite{nnnll}. Nonperturbative effects are purely of nonpotential nature 
and the leading ones are carried by the chromoelectric field which mediates the 
singlet-octet interaction. From (\ref{pnrqcd0}) 
the quarkonium  energy levels have been calculated up to  
$\als^5 \ln \mu^\prime$ and are given by\cite{nnnll} 
\begin{eqnarray}
E_{n,l,j}  &=&  \langle nl\vert { {\bf p}^2\over m} + V_s(\mu^\prime) + 
{V_s^{(1)}(\mu^\prime) \over  m}  
+ {V_s^{(2)}(\mu^\prime) \over  m^2 }  \vert nl \rangle \nn \\
& & \!\!\!\!\!\!\!  -i{g^2 \over 3 N_c}T_F  \int_0^\infty \!\! dt \,
\langle n,l |  {\bf r} e^{it( E_n - H_o)} {\bf r}  | n,l \rangle  
\langle {\bf E}^a (t) \phi(t,0)^{\rm adj}_{ab} {\bf E}^b (0)
 \rangle( \mu^\prime)
\label{nnnlospectrum}
\end{eqnarray}
where
$E_n = -  m  {C_F^2 \alpha_{\rm s}^2  \over 4  n^2}$, 
$H_o = {{\bf p}^2\over m } + V_o^{(0)}$  
and $ | n,l \rangle $ are the Coulomb wave functions. $V_s$  is taken
at three loop leading log and the 
other singlet potentials in the $1/m$ expansion
at the appropriate order in $\alpha_s$ for the requested accuracy.  
If $\lQ \ll mv^2$,  also the scale $m v^2$ can be integrated out perturbatively from
the above formula and, noticing  that:
$
m {\epsilon_n  n^6  \over (m C_F \als)^4} 
= {1 \over 3 N_c} T_F \left\langle n,l \left|  {\bf r} {1 \over E_n - H_o }
    {\bf r }  \right| n,l \right\rangle, 
$
the non-perturbative contributions reduce simply to the well-known
Voloshin--Leutwyler formula \cite{voloshin} 
\begin{equation}
\delta E^{\rm V-L}_{nl} = m {\epsilon_n  n^6  \pi^2  \langle F^2(0) \rangle
 \over (m C_F \als)^4},   
\label{contvole}
\end{equation}
($\epsilon_n$ is a (known) number of order 1)  and thus the only nonperturbative contribution is 
given in this case at leading order by  the local gluon condensate  
$\langle F^2(0)\rangle \equiv \langle (\als
/ \pi) F^a_{\mu\nu}(0) F^{a\,\mu\nu}(0) \rangle $.  This  is a well known nonperturbative 
parameter that can be calculated also inside QCD vacuum models \cite{rein,pol}.
Further nonperturbative corrections are still
 of nonpotential type and are suppressed.
For quarkonium of a typical size 
smaller than $1/\lQ$, the most relevant operator for the nonperturbative dynamics
 is the bilocal gluon condensate 
$\langle {\bf E}^a (t) \phi(t,0)^{\rm adj}_{ab} {\bf E}^b (0) \rangle $, 
which belongs to 
the class of th non-local nonperturbative  gluon condensate 
$\langle F^a_{\mu\nu} (x) \phi(x,0)^{\rm adj}_{ab} F^b_{\lambda\rho} (0) \rangle$.
Let me open a small parenthesis by summarizing
our knowledge of such operator and  what model independent information 
can be obtained by the EFT.\par
Different ~parametrizations have been ~proposed 
\cite{sv,bali,latpi,nonlocal} for the nonlocal gluon condensate. 
Because of its Lorentz structure, the correlator is in general
described by two form factors. A convenient choice of these
consists in the chromoelectric and chromomagnetic correlators:
$$
\langle {\bf E}^a (x) \phi(x,0)^{\rm adj}_{ab} {\bf E}^b (0) \rangle\, ,
\qquad\qquad
\langle {\bf B}^a (x) \phi(x,0)^{\rm adj}_{ab} {\bf B}^b (0) \rangle.
$$
The strength of the correlators is of the order of the 
gluon condensate. In the long range ($x^2 \to \infty$) they fall
off exponentially (in the Euclidean space) with some typical correlation lengths
that can be measured on the lattice \cite{bali,latpi}.
In \cite{bali} two different lengths have been obtained for the 
chromoelectric ($T_g^E$) and the chromomagnetic correlators ($T_g^B$):
\begin{eqnarray}
T_g^E \neq T_g^B \simeq 0.1\hbox{--}\,0.2 \hbox{~fm ~(quenched)}.
\label{latt3}
\end{eqnarray}
A recent sum-rule calculation \cite{hyb} gives $T_g^E < T_g^B$. 
The sum rule turns out not to be stable for the chromoelectric correlator, 
while for the chromomagnetic correlation length it gives  
\begin{equation}
{T_g}^B  = 0.11^{+0.04}_{-0.02} \hbox{~fm ~(quenched)}.
\label{sum2}
\end{equation}
The interesting thing is that these  
 $T_g^E$ and $T_g^B$,  which control the form of the nonperturbative correction
entering the energy levels, 
have a precise physical
interpretation. Their inverses correspond to the masses of the 
lowest-lying vector and pseudovector gluelumps\footnote{The gluelump is a state
formed by a $Q\bar{Q}$ pair in the adjoint representation plus a gluon, all
in the same spatial point. They are measured on the lattice and 
their masses are, roughly speaking,
 the limit of the hybrids static energies when the typical $Q\bar{Q}$ distance $r$ 
goes to zero.}, respectively. 
This can be explicitly seen in the short-range limit, $r=\vert{\bf x}\vert \to 0$, 
where the hybrids  operators can be 
explicitly constructed \cite{pnrqcd,hybrids}. The suitable effective field 
theory is pNRQCD in the static limit. Gluelump operators
are of the type ${\rm Tr}\{{\rm O}H\}$, where ${\rm O} = O^aT^a$ corresponds 
to a quark--antiquark state in the adjoint representation (the octet) and 
$H = H^a T^a$ is a gluonic operator. By matching the QCD static hybrid 
operators into pNRQCD, we get the static energies (also called potentials)
of the hybrids, which reduce to the gluelumps in the limit $r\to 0$.
 At leading order in the multipole expansion, they read
\begin{equation}
V_H(r) = V_o(r) + {1 \over T_g^H} + O(r^2);
\quad \langle H^a(t) \phi(t,0)^{\rm adj}_{ab}H^b(0)\rangle^{\rm non-pert.} 
\simeq h \, e^{- i t/T_g^H}. 
\label{potglue}
\end{equation}
%\begin{eqnarray}
%& & V_H(r) = V_o(r) + {1 \over T_g^H},
%\label{potglue}\\
%& & \langle H^a(t) \phi(t,0)^{\rm adj}_{ab}H^b(0)\rangle^{\rm non-pert.} 
%\simeq h \, e^{- i t/T_g^H}. 
%\nonumber
%\end{eqnarray}
Since hybrids are classified in QCD according to the representations of
$D_{\infty,h}$, while in pNRQCD, where we have integrated out the length $r$, 
their classification is done 
according to the representations of $O(3)\times C$, the static hybrid
short-range spectrum is expected to be more degenerate than the long-range one  
\cite{pnrqcd,hybrids}. The lattice measure of the hybrid potentials 
done in \cite{morn} confirms this feature. In \cite{pnrqcd} it has
been shown that the quantum numbers attribution of pNRQCD to the short-range
operators, and the expected $O(3) \times C$ symmetry of the effective field 
theory match the lattice measurements. By using only ${\bf E}$ and
${\bf B}$ fields and keeping only the lowest-dimensional representation 
we may identify the operator $H$ for the short-range hybrids called $\Sigma_g^{+\,'}$ 
(and $\Pi_g$) with ${\bf r}\cdot{\bf E}$ (and ${\bf r}\times{\bf E}$) 
and the operator $H$ for the short-range hybrids called $\Sigma_u^{-}$ 
(and $\Pi_u$) with ${\bf r}\cdot{\bf B}$ (and ${\bf r}\times{\bf B}$). 
Hence, the corresponding static energies for small $r$ are
$$
V_{\Sigma_g^{+\,'},\Pi_g}(r) = V_o(r) + {1 \over T_g^E}, \qquad\qquad
V_{\Sigma_u^{-},\Pi_u}(r) = V_o(r) + {1 \over T_g^B}.
$$
The lattice measure of \cite{morn} shows that, in the short range, 
$\!V_{\Sigma_g^{+\,'},\Pi_g}(r)\! >\! V_{\Sigma_u^{-},\Pi_u}(r)$. 
This supports the sum-rule prediction \cite{hyb} that the pseudovector 
hybrid lies lower than the vector one, i.e. $T_g^E < T_g^B$.\par
Summing up, the message is that, within the EFT,
 we can obtain quite a number of model-independent 
information on the nonperturbative 
contributions to the energy levels which are carried by 
these nonlocal condensates and we can establish interesting relations
 between these 
objects and physical 'nonperturbative' entities like the hybrids.
\par

Let us now briefly consider the situation when $mv >\lQ> mv^2$.
The infrared sensitivity of the quark--antiquark static potential at three loops \cite{pnrqcd}
 signals that 
it may become sensitive to non-perturbative effects if the next relevant scale 
after $m v$ is $\lQ$. Indeed, in the situation $mv \gg \lQ \gg mv^2$, the leading non-perturbative 
contribution (in $\als$ and in the multipole expansion) to the static potential reads \cite{pnrqcd}
\begin{equation}
V_0(r)^{\rm nonpert}  
= -i{g^2 \over N_c} T_F {r^2\over 3} \int_0^\infty \!\! dt \, e^{-itC_A \als /(2 r)} 
\langle {{\bf E}^a}(t) \phi(t,0)^{\rm adj}_{ab} {\bf E}^{b}(0) 
\rangle(\mu^\prime). 
\label{potnon}
\end{equation}
This term has to be summed to $V_0^{\rm pert}$ and it 
explicitly cancels  the dependence of the perturbative 
static potential on the infrared scale $\mu^\prime$. 
It is interesting to note that the leading 
contribution in the $\lQ/mv^2$ expansion of $V^{\rm non-pert}$ (obtained by
putting the exponential equal to 1) cancels the order $\lQ^3 r^2$ renormalon
that affects the static potential (the leading-order renormalon, of order $\lQ$, cancels against the 
pole mass). Therefore, also in the renormalon language, the above operator is the relevant 
non-perturbative contribution to the static potential in the considered kinematic situation.\par
The nonperturbative terms in the potential are in this case  organized in powers of 
$\alpha_s/(r\lQ)$, starting with a quadratic term, $r^2$. The actual form 
of these short range 
nonperturbative corrections to the potential depends on the actual form of the nonlocal 
correlator\cite{pnrqcd}. It is therefore very interesting to calculate this object inside 
models of the QCD vacuum, or with topological configurations on the lattice, also in relation 
to the recent claim of violation of the OPE expansion 'carried' by a nonperturbative short distance 
'string' term in the static singlet potential\cite{pol}.
\vskip -1truecm
\section{Nonrelativistic quark bound states  with a large characteristic \\  radius: 
  Wilson loops }
\vskip -0.15truecm
Fig. 1  shows that most of the quarkonia states lie in a region where the inverse of the size of the system 
is close to the scale $\lQ$. In this situation the potential can no longer be expressed as an 
expansion in $\alpha_s$.
 The  non-perturbative dynamics is already switched on at the binding (potential) scale 
and it is  contained in more extended objects than 
(local or non-local) gluon condensates: Wilson loops and
chromoelectric and chromomagnetic   field insertions on the Wilson loops.
Such operators can be eventually 
calculated on the lattice \cite{lattice}  or in QCD vacuum models 
\cite{vacuum,bak}. 
In \cite{aprot} the matching of QCD to pNRQCD has been performed at order $1/m^2$ 
in the general situation $\lQ \siml mv$. This has been proved 
to be equivalent to compute the heavy quarkonium potential at order $1/m^2$. 
More precisely, a pure potential picture emerges at the leading order in the ultrasoft 
expansion under the condition that all the gluonic excitations (hybrids) have a gap of $O(\lQ)$
with respect to the singlet. Such situation is confirmed by lattice simulations\cite{lattice}. 
Then if we  switch off the light fermions (pure
gluodynamics), only the singlet survives and pNRQCD reduces to a pure two-particle 
NR quantum-mechanical system. Therefore, the situation {\it assumed} by all potential models, 
may be  now rigorously {\it derived} under a specific set of 
circumstances (for more details see\cite{aprot}).
In this situation, pNRQCD involves 
only  a bilinear colour singlet field, $S({\bf x}_1,{\bf x}_2,t)$, 
\be
{\cal L}_{\rm pNRQCD} = S^\dagger 
\bigg( i\partial_0 -h_s({\bf x}_1,{\bf x}_2, {\bf p}_1, {\bf p}_2)\bigg) S, 
\label{pnrqcdl}
\ee
where $h_s$ is the Hamiltonian of the singlet, ${\bf p}_1= -i \bfnabla_{{\bf x}_1}$ 
and ${\bf p}_2= -i \bfnabla_{{\bf x}_2}$. It has the following expansion up to order $1/m^2$ 
\begin{equation}
h_s  = 
{{\bf p}^2_1\over 2 m_1} +{{\bf p}^2_2\over 2 m_2} + V^{(0)}
+{V^{(1,0)} \over m_1}+{V^{(0,1)} \over m_2}+ {V^{(2,0)} \over m_1^2}
+ {V^{(0,2)}\over m_2^2}+{V^{(1,1)} \over m_1m_2}.
\label{hss}
\end{equation}
Higher order effects in the $1/m$ expansion as well as extra ultrasoft degrees of freedom\cite{pnrqcd,aprot,glue}, 
such as hybrids and pions  can be systematically included and may eventually affect the leading 
potential picture (like in the perturbative regime ultrasoft gluons \cite{pnrqcd,aprot}).
We shall use the following notations: $\langle \dots \rangle$ will stand for the average 
over the Yang--Mills action, $W_\Box$ for the rectangular static Wilson loop of dimensions 
$r\times T$ and $\langle\!\langle \dots \rangle\!\rangle 
\equiv \langle \dots W_\Box\rangle / \langle  W_\Box\rangle$. 
We define $\lla O_1(t_1)...O_n(t_n)\rra_c$ 
as the {\it connected} Wilson loop. 
We get in terms of Wilson 
loops \cite{aprot}
\be 
V^{(0)}(r) = \lim_{T\to\infty}{i\over T} \ln \langle W_\Box \rangle, 
\label{v0}
\ee
\begin{equation}
V^{(1,0)}(r)=
-{1 \over 2} \lim_{T\rightarrow \infty}\int_0^{T}dt \, t \, \lla 
g{\bf E}_1(t, {\bf x}_1)\cdot g{\bf E}_1(0, {\bf x}_1) \rra_c 
= V^{(0,1)}(r).
\label{Em12}
\end{equation}
All the obtained expression for the potentials are
manifestly gauge invariant and are  also correct at any power in $\als$
in the perturbative regime.
For the terms of order $1/m^2$ we have the general decomposition 
\begin{equation} 
V^{(2,0)} = {1 \over 2}\left\{{\bf p}_1^2,V_{{\bf p}^2}^{(2,0)}(r)\right\}
+{V_{{\bf L}^2}^{(2,0)}(r)\over r^2}{\bf L}_1^2 
+ V_r^{(2,0)}(r) + V^{(2,0)}_{LS}(r){\bf L}_1\cdot{\bf S}_1,
\label{mm2}
\end{equation}
where ${\bf L}_j \equiv {\bf r} \times {\bf p}_j$. 
Similar definitions hold  for $V^{(0,2)}$ and $V^{(1,1)}$.
All the potentials (i.e. the functions of $r$)
 in $V^{(2,0)}, V^{(0,2)}, V^{(1,1)}$ are obtained factorized in two
contributions. The first one contains the hard physics at the scale $m$ which is calculated 
as a series expansion in $\alpha_s$ at the appropriate hard 
scale\cite{aprot,nrrev}. The second contribution 
at the scale $1/r$ ,
enjoys a closed expression in terms of average values
of two or more 
chromoelectric and chromomagnetic insertions in the presence 
of the Wilson loop \cite{aprot},
similarly to ({\ref{v0}),(\ref{Em12}).
The explicit expressions can be found in \cite{aprot}.
{\it Therefore, the calculation of the dynamics, i.e. the calculation of the potential, is 
in this way reduced to a calculation in 
pure gluodynamics  at the lowest scale.}
On one hand, such objects are very conveniently evaluated on the lattice since they are pure glue 
objects and contain only one scale.
On the other hand, such objects can be very well evaluated using nontrivial topological configurations 
inside a model of the QCD vacuum \cite{vacuum,bak}. 
We emphasize that,
 since the potential we get here is a well defined quantity, derived from QCD via a systematic 
and unambiguous procedure, and {\it complete} up to order $1/m^2$, 
it is not affected by the usual ambiguities (ordering, retardation corrections, etc.), which affect 
all potential models and all phenomenological reductions of Bethe--Salpeter kernels \cite{rev}. 
For the same reason the above results are the  relevant ones for the study of the properties 
of the QCD vacuum in the presence of heavy sources. 
The average value of the field insertions
in the presence of the Wilson loop contain all the relevant information about the heavy quark dynamics.
This is much more detailed and quantitative with respect to the generic request that topological 
configuration should originate an area law behaviour into the Wilson loop. 
If we calculate these field insertions on the Wilson loop using some dominant topological 
configurations we have then  a simple procedure to relate the result,
 on one hand to  standard lattice 
simulations of the same objects and, on the 
other hand, directly to the spectrum. This last feature 
descends from the fact that within the EFT we are able to 
fully take into account the perturbative 
contributions  and we have a clear power counting that selects 
the terms contributing at a given order in the $v$ expansion.\par
%Moreover, the above non-perturbative 
%formulation of pNRQCD has translated this problem into obtaining 
%the power counting of the different potentials. This is expected to be of some 
%advantage. Since the scale $mv$ has been integrated out, 
%the power counting of pNRQCD is simpler and maybe different 
%from the one of NRQCD\footnote{This new power counting has been proposed 
%in \cite{aprot}
%and recently applied to charmonium production in\cite{leib}.} 

\section{Topology and heavy quark bound states}

The outlined study is rigorous and allows a systematic disentanglement
of the high from the low energy scales of the heavy quarkonium system under
study. In this specific sense also perturbative and nonperturbative effects 
turn out to be disentangled. 
Thus the nonrelativistic dynamics 
of heavy quark bound states is surely an adequate 
benchmark to test ideas and models of QCD topology and QCD vacuum structure. \par
As I have shown, when $mv \sim \lQ$, then the whole nonperturbative 
$Q\bar{Q}$ interaction at order 
$1/m^2$ turns out to be expressed in terms of average values of Wilson loops
and field insertions into the Wilson loop.  We know that 
in the confined phase the Wilson loop is dominated  by an area law\cite{rev} which is 
related to the formation of an interquark confining  flux tube\cite{flux}. 
Such feature has been obtained using different 
topological configurations: gluon Abelian projected fields\cite{pol},
monopoles\cite{pol}, center vortices\cite{fab}. In all these cases
an approximated area law is obtained and it is debated what are the relevant 
topological configurations that really drive the confinement dynamics.
Many lattice investigations have been performed and have ended up with the 
notion that in a way all these topological configurations are related\cite{latinv}.
Several QCD vacuum models  exist in the literature\cite{sv,bak,vacuum,rev}
and all of them obtain an area law for the Wilson loop.
However, there are quite a number of differences between different vacuum models 
as well as between different topological configurations, but 
such differences are not apparent 
at the level of the static Wilson loop. 
In this case, typically we need only one parameter, the string tension $\sigma$, to encode the 
information of a constant energy density in the confining flux tube. 
However, there are many ways in which this can be realized in relation to 
the actual profile of the flux tube. It is well known for example that 
while the $Q\bar{Q}$ flux tube  measured on the lattice in QCD is quite fat,
the one measured on the lattice in the Abelian projection is very thin\cite{bali2}.
\par
On the other hand in order to describe 
the nonperturbative heavy quark dynamics at order $1/m^2$ we need not only 
the Wilson loop area law 
behaviour but also the behaviour of the chromoelectric and 
chromomagnetic field insertion into the Wilson loop.  These last contain information on the profile of the flux tube and turn out to be much more sensitive 
to the QCD vacuum models or the topological configurations\cite{rev,vacuum}
\footnote{In two models for the QCD vacuum,
stochastic vacuum model\cite{sv} and Dual QCD\cite{bak}, 
the profile of the flux tube is controlled by the correlation length $T_g$ given in equation (\ref{latt3}) and thus it is ultimately related 
to hybrids configurations.} .
 Thus, it is much more interesting to study the full 
non relativistic dynamics of heavy quarks in place of the much more qualitative 
statement about finding out confining configurations into the Wilson loop. 
In particular, it would be interesting to have a lattice 
calculations of (\ref{v0}), (\ref{Em12}), (\ref{mm2}) and the other 
 $1/m^2$ potentials contained in \cite{aprot},
 using monopoles configurations\cite{pol} versus  center vortices\cite{rein,fab}. 
In conclusion, I would like to invite the people working in the field to calculate 
local gluon condensates, nonlocal gluon condensates, Wilson loops and field 
insertions into the Wilson loop in their topological 
models and to relate them to the phenomenological
data using the unified and systematic frame of the EFT that I presented here.
\vskip -0.6truecm
\section*{Acknowledgments}
It is my great pleasure to  thank the Organizers of this School,
P. Bicudo, P. Sacramento,  J. Seixas, 
V. Rocha Vieira and in particular Emilio Ribeiro, for the invitation and 
for providing  such an 
interesting programme and a very nice, stimulating and friendly atmosphere.
I thank A. Vairo for reading the manuscript and discussions. Many discussions 
with P. Bicudo and E. Ribeiro are gratefully acknowledged.

\end{document}